\documentclass[twocolumn,aps,prl,showpacs,superscriptaddress]{revtex4}
\usepackage[dvips]{graphicx}
\usepackage{bm}
\begin{document}
\title{Exact time correlation functions for $N$ classical
 Heisenberg spins in the `squashed' equivalent neighbor model}
\author{Marco Ameduri}
\email{ma22@cornell.edu}  
\affiliation{Weill Cornell Medical College in Qatar, Qatar Foundation -
 Education City, P. O. Box 24144, Doha, Qatar}
\affiliation{Max-Planck-Institut f{\"u}r Physik komplexer Systeme,
N{\"o}thnitzer
 Stra{\ss}e 38, D-01187 Dresden, Germany}
\author{Richard A. Klemm}
\email{rklemm@mpipks-dresden.mpg.de} 
\affiliation{Max-Planck-Institut f{\"u}r Physik komplexer Systeme,
N{\"o}thnitzer
 Stra{\ss}e 38, D-01187 Dresden, Germany}
\date{\today}
\begin{abstract}
 We present exact integral representations of the time-dependent spin-spin
 correlation functions for the classical Heisenberg $N$-spin  `squashed' equivalent
 neighbor model,  in which  one spin is coupled via the Heisenberg
 exchange interaction with strength $J_1$ to the other $N-1$ spins, each of which is coupled
 via the Heisenberg exchange interaction with strength $J_2$ to the remaining $N-2$
 spins.  At low temperature $T$ we find that the $N$ spins 
 oscillate in four modes, one of which is a central peak for a
 semi-infinite range of the values of the exchange
 coupling ratio.   For the $N=4$ case of four spins on a squashed 
tetrahedron, detailed
 numerical evaluations of these results are presented.  As
 $T\rightarrow\infty$,  we calculate
 exactly the long-time asymptotic behavior of the correlation
 functions for arbitrary $N$, and compare our results with those obtained for
 three spins on an isosceles triangle.
\end{abstract}
\vskip0pt\vskip0pt
\maketitle


\section{I. Introduction}
 Recently there has been a growing interest in the study of the
 properties of magnetic molecules.
 \cite{Mn12,Cr3tetra:1,Cr3tetra:2,Fe3tetra:1,Fe3tetra:2}
 The defining characteristic
 of these substances is the presence of a small cluster of magnetic
 ions located at the center of each molecule and surrounded by
 a complicated structure of non-magnetic chemical
 ligand groups. In general, the strength of the magnetic interaction
 between ions located in different molecules is negligible in
 comparison to the strength of their intramolecular interactions.
 Therefore, measurements of the magnetic properties of macroscopic
 samples reflect the underlying magnetic interactions within a
 single molecule.

 The list of synthesized magnetic molecules has been constantly
 growing, even though most of the experimental activity has been
 focused on the determination of the magnetic properties of 
 a molecule containing twelve manganese ions at its core, often
 referred to as Mn$_{12}$. The theoretical tools
 currently used to describe the behavior of this relatively
 complicated structure are still rudimentary, and are based on a
 single-spin phenomenological Hamiltonian. \cite{Mn12} 
It is important to notice
 that a number of molecular structures containing smaller numbers
 of magnetic ions have already been synthesized. For some of these
 structures it is possible to perform a more detailed theoretical
 analysis of their magnetic behavior starting from a many-spin
 Hamiltonian.

 Among the smaller clusters are a regular tetrahedron of ${\rm
 Cr}^{3+}$ ions ($S=\frac{3}{2}$) \cite{Cr3tetra:1,Cr3tetra:2}, Cr$_4$, and
 a squashed tetrahedron of ${\rm Fe}^{3+}$ ions ($S=\frac{5}{2}$)
 \cite{Fe3tetra:1,Fe3tetra:2}, Fe$_4$. For increasing values of the spin
 a description in terms of classical spins is expected to capture
 many of the features of the system \cite{qm:to:class:1,qm:to:class:2}.
 In the present paper we provide exact expressions for the time-dependent spin-spin
 correlation functions for the classical Heisenberg 
$N$-spin squashed equivalent neighbor model,
 which is the $N$-spin generalization of four classical Heisenberg
 spins on the corners of a squashed tetrahedron. Specific numerical results for
 the squashed tetrahedron case, $N=4$, are provided.  Analogous
 studies have recently appeared for three spins on an isosceles
 triangle and on a chain \cite{isosceles}, for four spins on a square
 ring \cite{square}, and for the equivalent neighbor model of $N$
 classical spins \cite{eq_neighbor}. Quantum time-dependent
 correlation functions have been computed for a dimer
 \cite{qm:to:class:1} and for three spins on an equilateral triangle
 \cite{qm:to:class:2}, and for a dimer of classical and quantum spins
 in a constant magnetic field \cite{dimer:field}. The availability of
 time-dependent correlation functions is necessary to analyze
 neutron scattering experiments.

 In Section II we define the Hamiltonian system to be
 studied and write the corresponding partition function. In Section
 III  we present the constraints upon the various correlation
 functions. In Section
 IV we present our analytic results
 for arbitrary $N$.  We evaluate the long-time behavior of
 the correlation functions at infinite temperature $T$, and provide
 analytic formulae
 for the low-$T$ modes for arbitrary $N$. In Section V, we present numerical
 results at low $T$ for the squashed tetrahedron, $N=4$.  Section VI
 contains our conclusions. A collection of intermediate steps
 useful to the calculations is compiled in the Appendix.


\section{II. The model} 

 We consider $N$ classical spins of unit magnitude, $|{\bm S}_{i}|=1$,
 interacting according to the Hamiltonian
\begin{equation} \label{H1}
 H = -{J_2\over{2}}\sum_{{i,j=1}\atop{i\ne j}}^{M}{\bm{S}}_i\cdot{\bm{S}}_j -J_{1}{\bm{S}}_{N}\cdot\sum_{i=1}^{M}{\bm{S}}_i,
\end{equation}
where
\begin{equation}
M\equiv N-1\ge2.\label{Nstar}
\end{equation}

 Introducing the total spin ${\bm S}=\sum_{i=1}^{N}{\bm S}_{i}$ and the
 auxiliary variable ${\bm S}_{1\rightarrow M}=\sum_{i=1}^{M}{\bm
 S}_{i}={\bm S}-{\bm S}_N$, the
 Hamiltonian (\ref{H1}) can be written as
\begin{equation} \label{H2}
 H = -\frac{J_{1}}{2} {\bm S}^{2}
     -\frac{J_{2}-J_{1}}{2}{\bm S}_{1\rightarrow M}^{2},
\end{equation}
 where we have dropped the constant energy $(J_{1}+MJ_{2})/2$.
 The partition function can then be calculated following the technique
 described in \cite{square,ciftja:squashed}. Letting $s=|\bm{S}|$ and
 $x=|{\bm{S}}_{1\rightarrow M}|$, one obtains
\begin{eqnarray} \label{Z}
 Z& =& \int_0^{M}dx{\cal D}_{M}(x)\int_{|x-1|}^{x+1}s ds\exp(-\beta H)\\
&=& {{ e^{\alpha}}\over{\alpha}} \int_{0}^{M} dx
     {\cal D}_{M}(x) \exp(\alpha\gamma x^{2}) \sinh(2\alpha x),
\end{eqnarray}
 where $\beta = (k_{{\rm B}}T)^{-1}$,
 $\alpha=\beta J_{1}/2$,  $\gamma =J_{2}/J_{1}$, and  ${\cal D}_M(x)$ is the
 classical $M$-spin density of states,\cite{eq_neighbor} which we redisplayed
 in Eq. (\ref{DOS}) in the Appendix.

 In order to compute the time-dependent correlation functions, we
 first solve the classical equations of motion appropriate for the
 Hamiltonian, Eq. (\ref{H2}),
\begin{equation}
  \dot{\bm{S}}_{N,1\rightarrow M} = J_{1} {\bm{S}}_{N,1\rightarrow M} \times {\bm{S}}
\end{equation}
and $\dot{\bm{S}}=0$, so that ${\bm{S}}$ is a constant of the motion.
 Following the technique illustrated in \cite{isosceles,square,eq_neighbor}, we
 obtain
\begin{eqnarray}
 {\bm S}_{N,1\rightarrow M}(t)& =& C_{N,1\rightarrow M} \hat{\bm{s}} +
    A_{N,1\rightarrow M}\nonumber\\
  & &\times\left[ \cos(st^{*})\hat{\bm{x}}-\sin(st^{*})\hat{\bm{y}}\right],
\label{SNoft}
\end{eqnarray}
 where $t^{*}=J_{1}t$, $\hat{\bm{s}}={\bm{S}}/s=
 \hat{\bm{x}} \times \hat{\bm{y}}$, $C_{N}=(s^{2}-x^{2}+1)/(2s)$,
 $C_{1\rightarrow M}=(s^{2}+x^{2}-1)/(2s)$, $A_{N}^{2}=1-C_{N}^{2}$, and
 $A_{N}=-A_{1\rightarrow M}$. 

 We must also consider the equations of motion for the ${\bm S}_{i}(t)$,
$i=1, 2,\ldots, M$.  In order to calculate the time correlation functions, symmetry
allows us to choose just one of them,  $i=1$.  We then write ${\bm
S}_{1\rightarrow M}={\bm S}_1+{\bm S}_{2\rightarrow M}$,
and solve
\begin{equation}
\dot{\bm{S}}_{1,2\rightarrow M}=J_2{\bm S}_{1,2\rightarrow M}\times{\bm
S}+(J_1-J_2){\bm S}_{1,2\rightarrow M}\times{\bm S}_N.
\end{equation}

After defining $S_{1\pm}=S_{1x}\pm i S_{1y}$,
 we obtain, 
\begin{eqnarray}
  S_{1\pm}(t)& =& -\frac{A_{N}S_{1z0}}{C_{1\rightarrow M}} \exp(\mp i st^{*})
       - \frac{A_{N}\Delta S_{1z0}}{2(C_{1\rightarrow M}\mp x)}\nonumber\\
& &\times \exp \left\{
       i \left[ \mp s+(1-\gamma)x \right] t^{*} + i \phi_{0}
       \right\} \nonumber \\
      & & - \frac{A_{N}\Delta S_{1z0}}{2(C_{1\rightarrow M}\pm
x)}\nonumber\\
& &\times \exp \left\{
       i \left[ \mp s - (1-\gamma)x \right] t^{*} - i \phi_{0}
       \right\}, \label{Sipm} \\ 
  S_{1z}(t)& =& S_{1z0} + \Delta S_{1z0}
       \cos \left[ (1-\gamma)xt^{*} + \phi_{0} \right],
       \label{Si3}
\end{eqnarray}
where $\phi_0$ is an arbitrary phase, and similar equations for the 
components of ${\bm{S}}_{2\rightarrow M}$.
 After combining these equations with analogous ones for the components of
 ${\bm{S}}_{1\rightarrow M}$, 
 the constants appearing in Eqs. (\ref{Sipm}) and (\ref{Si3}) must satisfy
\begin{eqnarray}
 S_{1z0} &= &\frac{C_{1\rightarrow M}}{2} \left( 1+\frac{1-y^{2}}{x^{2}}
            \right), \label{S1z0}\\
 (\Delta S_{1z0})^{2} &= &\frac{A_{1\rightarrow M}^{2}}{x^{2}} \left[ 1-
            \frac{(x^{2}-y^{2}+1)^{2}}{4x^{2}} \right],\label{DeltaS1z0}
\end{eqnarray}
 where $y=|{\bm{S}}_{2\rightarrow M}|$.  

Previously, we solved these equations
 for the simplest case, $M=2$, for which $y=1$ is not a variable.\cite{isosceles}  In that
 case, the correlation functions were obtained from the double integrals over
 $x$ and $s$, according to the weighting factors in Eq. (\ref{Z}).  
For $M\ge3$,
 however, $y$ can vary over the entire range $0\le y\le M-1$.   
 Hence, for the explicit evaluation of the correlation functions with $M\ge3$,
 it is
 useful to rewrite the expression of the partition function (\ref{Z})
 in terms of a triple integral over $s$, $x$, and $y$,
\begin{equation}\label{Z3}
 Z=\int_{0}^{M-1}{\cal D}_{M-1}(y) dy \int_{|y-1|}^{y+1} dx
   \int_{|x-1|}^{x+1} s ds \exp (-\beta H).\label{Z2}
\end{equation}


\section{III. Constraints}

 In this section we analyze the constraints upon
  the time-dependent spin-spin correlation functions
\begin{equation} \label{Cij}
 {\cal C}_{ij}(t)= \langle {\bm S}_{i}(t) \cdot {\bm S}_{j}(0) \rangle,
\end{equation}
 where the thermal average $\langle \ldots \rangle$ is  performed by averaging
over the arbitrary phase $\phi_0$ and the variables $s$, $x$, and $y$, with
 respect to the canonical ensemble defined by Eq. (\ref{Z3}).
 Due to  the symmetry of the molecule, only four of the $N(N+1)/2$ 
correlation functions
in Eq. (\ref{Cij}) are distinct. 
 We write these as ${\cal C}_{11}(t)$, ${\cal C}_{12}(t)$, ${\cal C}_{1N}(t)$, and ${\cal C}_{NN}(t)$.
 Conservation of
 the total spin adds a  constraint,
\begin{eqnarray}
 \langle s^{2} \rangle &=& {\cal C}_{NN}(t) + M{\cal C}_{11}(t)+
    2M{\cal C}_{1N}(t)\nonumber\\& & + M(M-1){\cal C}_{12}(t).\label{s2}
\end{eqnarray}
 Finally, by writing the multispin correlation function $\langle {\bm S}_{1\rightarrow M}(t)\cdot {\bm
 S}_{1\rightarrow M}(0)\rangle$ in two ways, we
 find a second constraint between two of the
 correlation functions,
\begin{equation}\label{C14}
\langle sC_N\rangle={\cal C}_{NN}(t)+M{\cal C}_{1N}(t),\label{sCN}
\end{equation}
where the constant $C_N$ is given just below Eq. (\ref{SNoft}).
 The two remaining independent correlation functions ${\cal C}_{NN}(t)$ and ${\cal C}_{11}(t)$ must
 then be calculated by explicitly substituting into
 Eq. (\ref{Cij}) the time dependences obtained in Section II.
For ${\cal C}_{NN}(t)$, this is relatively simple, as one can just use Eq. (\ref{SNoft})
 for ${\bm S}_N(t)$, which is independent of $y$, to evaluate it.  This leads
 to 
\begin{equation}
{\cal C}_{NN}(t)=\langle C_N^2+A_N^2\cos(st^*)\rangle,\label{CNN}
\end{equation}
which can be evaluated using the simplified weighting factors present in
 Eq. (\ref{Z}).  From
 Eq. (\ref{C14}), this simplification
 also applies for ${\cal C}_{1N}(t)$. We note that Eq. (\ref{CNN}) differs
 from the expression for the autocorrelation function in the $N$-spin
 classical Heisenberg equivalent neighbor model only by the $x$ dependence of
 the Hamiltonian,\cite{eq_neighbor}
 which is irrelevant as $T\rightarrow\infty$.

The challenge is to calculate ${\cal C}_{11}(t)$. It is
 useful to separate the expression for ${\cal C}_{11}(t)$ into the
 four integrals $I_{i}(t)$ ($i=0,\ldots,3)$,
\begin{equation} \label{C11}
 {\cal C}_{11}(t) = \sum_{i=0}^{3} I_{i}(t).
\end{equation}
 The explicit triple integral representations of the 
 $I_{i}(t)$ valid for arbitrary $T$ are given in the Appendix, where
 it is also shown how to reduce
 them to double integrals. 


\section{IV. Analytic Results for Arbitrary $N$}

\subsection{A. Infinite temperature limit}

 Here we present our results for the correlation functions with general
 $N$ values as $T\rightarrow \infty$. 
 As shown in the Appendix, in the limit $T \rightarrow
 \infty$, the triple  integrals appearing in (\ref{C11}) can be reduced to
 single integrals.  For $N=4$, the relevant density of states appearing in Eq. (\ref{Z3})
 is ${\cal
 D}_2(x)={1\over{2}}\Theta(x)\Theta(2-x)$, so this reduction in the number of
 integrals is relatively simple. 
As $T\rightarrow\infty$, the different couplings appearing in the Hamiltonian
 become irrelevant for ${\cal C}_{NN}(t)$, so that it becomes equivalent to
 that of the $N$-spin equivalent-neighbor model,\cite{eq_neighbor}
\begin{equation}
\lim_{T\rightarrow \infty} {\cal C}_{NN}(t) = 1/N + M[\delta_N+f_N(t)],
\label{CNNinfT}\\
\end{equation}
where $f_N(t)\sim (t^*)^{-N}$ for $t^*\gg1$.
Since as $T\rightarrow\infty$, $\langle s^2\rangle = N$, $\langle x^2\rangle =
M$, and $\langle y^2\rangle = M-1$, from Eqs.  (\ref{sCN}) and 
 (\ref{CNNinfT}), 
 we have
\begin{eqnarray}
\lim_{T\rightarrow \infty} {\cal C}_{1N}(t)& =& 1/N-\delta_N-f_N(t).
\end{eqnarray}

For ${\cal C}_{11}(t)$ and ${\cal C}_{12}(t)$, even as $T\rightarrow\infty$, the situation is more
 complicated, as the results depend crucially upon the values of 
$\gamma=J_2/J_1$.
 As $t\rightarrow \infty$, the time-dependent trigonometric functions
 in $I_{1}, I_{2}$, and $I_{3}$  oscillate increasingly rapidly and
 yield  vanishing contributions to
 $\lim_{t\rightarrow\infty}{\cal C}_{11}(t)$, as stated by the Riemann-Lebesgue
 lemma \cite{lemma}. Therefore, for arbitrary $N$,
\begin{equation}
 \lim_{t\rightarrow \infty} {\cal C}_{11}^{\gamma \neq 1}(t) = I_{0}=\langle
S^2_{1z0}\rangle,\\
\end{equation}
We note that $I_0$ depends upon $N$, and is a rather messy triple integral,
 but that as $T\rightarrow\infty$, can be evaluated exactly, as shown in the Appendix.

 At infinite temperature one obtains for $N=4$,
\begin{eqnarray} 
 \lim_{{t\rightarrow \infty}\atop{T\rightarrow \infty}}
     {\cal C}_{44}(t) = \frac{1}{4} + 3\delta_{4}\approx 0.436345,\label{C44inft}
\end{eqnarray}
 where  $\delta_{4} = -(11/180) + (8/45)\ln 2 \approx
 0.062115$, \cite{square,eq_neighbor} and
\begin{eqnarray} 
 \lim_{{t\rightarrow \infty}\atop{T\rightarrow \infty}}
     {\cal C}_{11}^{\gamma \neq 1}(t)& \approx& 0.355496,\label{approx_limit}
\end{eqnarray}
 the exact expression for which is given in (\ref{exact_limit}) in the
 Appendix. In Table I in the Appendix, we also list the $T\rightarrow\infty$ values of
 $\lim_{t\rightarrow\infty}{\cal C}_{11}^{\gamma\ne1}(t)$ for $3\le N\le11$,
 and compare them with the $T\rightarrow\infty$ values of
 $\lim_{t\rightarrow\infty}{\cal C}_{NN}(t)$.  
We note that as $T\rightarrow\infty$, for each of these $N$ values, $\lim_{t\rightarrow\infty}{\cal
C}^{\gamma\ne1}_{11}(t)<\lim_{t\rightarrow\infty}{\cal
C}_{NN}(t)$.  As $T\rightarrow\infty$, $\lim_{t\rightarrow\infty}{\cal
C}_{NN}(t)$ decreases monotonically with increasing $N$ to $\frac{1}{3}$ as
$N\rightarrow\infty$.\cite{eq_neighbor}  Since  $\lim_{t\rightarrow\infty}{\cal
C}^{\gamma\ne1}_{11}(t)$ also decreases monotonically with increasing $N$,
and for $8\le N\le11$, its value is less than $\frac{1}{3}$, it appears
 that this
inequality is likely to hold for all $N$ values.

We now turn to the long-time asymptotic behavior of ${\cal C}_{11}(t)$
 at infinite $T$.
 Following the method described in \cite{isosceles}, we  first 
 define  $\delta{\cal 
 C}_{ij}(t)\equiv{\cal C}_{ij}(t)-\lim_{t\rightarrow\infty}{\cal C}_{ij}(t)$.
 For $\gamma=0$, the dominant behavior of $\lim_{T\rightarrow\infty}{\cal
 C}_{11}(t)$ is given by $I_3(t)$, but for $0\ne\gamma\ne 1$, it is given by
 $I_2(t)$.  At long times, $\overline{t}\gg1$, where
 $\overline{t}=(1-\gamma)t^*$, one can evaluate the asymptotic behavior as
 $T\rightarrow\infty$ exactly.   By integration by parts $M$ times, we find,
\begin{eqnarray}
\lim_{{T \rightarrow \infty} \atop{\overline{t} \gg 1}} \delta 
     {\cal C}_{11}^{\gamma \neq 0,1}(t) &\sim
     &\sum_{p=0}^{E(M/2)}\frac{A_{Mp}}{(\overline{t})^M}
f(M-2p)\nonumber\\
& &\times\cos[(M-2p)\overline{t}+M\pi/2],\label{C11longtime}
\end{eqnarray}
where
\begin{eqnarray}
f(y)&=&1+y^{-2}-\frac{(y^2-1)^2}{4y^3}\ln\Bigl(\frac{y+1}{y-1}\Bigr)^2
\label{fofy}
\end{eqnarray}
and $A_{Mp}$ is given in the Appendix.   Although the function $f(y)$ is 
non-analytic at $y=1$, it
 can be shown that its derivatives do not contribute to the long-time
 asymptotic behavior. 
In addition, for $t^*\gg1$, one can easily obtain the
 asymptotic expression of $I_3(t)$, leading to 
\begin{eqnarray}
\lim_{{T \rightarrow \infty} \atop{t^{*} \gg 1}} \delta
     {\cal C}_{11}^{\gamma=0}(t)& \sim
     &\frac{\sin(t^*)}{4t^*}\int_0^{M-1}dy{\cal D}_{M-1}(y)\nonumber\\
& &\times y^3f(y).\label{C11oftMgamma0}
\end{eqnarray}

In particular, for $N=4$, we obtain
\begin{eqnarray}
  \lim_{{T \rightarrow \infty}\atop{t^{*} \gg 1}} \delta
     {\cal C}_{44}(t)& \sim & - \frac{3}{4(t^{*})^{4}}
     \left[ \frac{3}{4} - \cos (4t^{*}) \right],\label{C44oft} \\
  \lim_{{T \rightarrow \infty} \atop{\overline{t} \gg 1}} \delta 
     {\cal C}_{11}^{\gamma \neq 0,1}(t) &\sim & -\frac{1}{8(\bar{t})^{3}}
     \Bigl[f(1)\sin (\bar{t})+f(3)\sin (3\bar{t})\Bigr],
\label{C11oftgammane01}\nonumber\\
& & \\
  \lim_{{T \rightarrow \infty} \atop{t^{*} \gg 1}} \delta
     {\cal C}_{11}^{\gamma=0}(t)& \sim &\left( \frac{23}{30} -
     \frac{9}{40} \ln 3 \right) \frac{\sin (t^{*})}{t^{*}},\label{C11oftgamma0}
\end{eqnarray}
where Eq. (\ref{C44oft}) was given previously.\cite{square,eq_neighbor}

 It is interesting to compare the present results to the analogous
 ones obtained for the isosceles triangle of spins, $N=3$. \cite{isosceles}
 For $N\ge4$, the
 infinite-$T$, long-time behavior of ${\cal C}_{11}(t)$ for $\gamma
 \neq 0,1$ is determined by the integral $I_{2}(t)$ given by Eq. 
 (\ref{I2}). For $N=3$, an additional contribution to the infinite-$T$,
 long-time behavior of ${\cal C}_{11}(t)$ arises from
 $I_3(t)$ given by Eq. (\ref{I3}). \cite{isosceles} For $N\ge3$, the correlation function ${\cal C}_{11}(t)$ for
 $\gamma\ne0,1$  decays
 slower than ${\cal C}_{NN}(t)$ [denoted ${\cal C}_{22}(t)$ in
 Ref. \cite{isosceles} for $N=3$], approaching its long-time 
asymptotic value at
 infinite temperature as $(t^*)^{-M}$. In the limiting situation $\gamma=0$,
 corresponding for $N=3$ to the three-spin chain (or `two-pronged star') and for $N\ge4$ to 
 an $M$-pronged star of spins equally coupled to a central one,
 as $T\rightarrow\infty$ and $t^*\gg1$,  the correlation function is dominated
 by $I_3(t)$.  In this case, ${\cal
 C}_{11}(t)$ approaches its asymptotic limit much more slowly, as
 $(t^{*})^{-1}$, as shown in Eq. (\ref{C11oftMgamma0}).


\subsection{B. Low-temperature correlation functions} \label{sec:lowT}

 At any finite temperature, it is not possible to reduce the
 time-dependent  correlation functions to a single integral
 representation, even for $N=4$. Since  the time-dependence of the integrand is a
 simple trigonometric  function, it is convenient to compute the
 Fourier transforms  of the ${\delta \cal C}_{ij}(t)$, quantities which
 are anyhow of direct experimental relevance in neutron scattering
 experiments. In this case, it is then possible to express the
 Fourier transforms in terms of a single integral representation,
 which then allows a precise and fast numerical integration.  We limit our
 numerical work to the case of the squashed tetrahedron, $N=4$ ($M=3$).

 We define the Fourier transform as usual as
\begin{equation}
 \delta \tilde{{\cal C}}_{ij} (\omega) = \frac{|J_{1}|}{\pi}
 \int_{-\infty}^{+\infty} dt \exp (i\omega t) \delta
 {\cal C}_{ij}(t).
\end{equation}

 The position of the various peaks
 as a function of $\gamma$ may be obtained analytically in the
 $T \rightarrow \infty$ limit through an asymptotic evaluation of
 the integrals, or numerically by plotting the curves at  large enough 
 values of
 $|\alpha| \propto 1/T$.  In the Appendix, we have sketched the
 derivation of the low-temperature mode frequencies for general $N$,  for both
 FM and AFM cases.  
For ferromagnetic couplings, we then find, 
\begin{eqnarray}
  \Omega_{1}(\gamma)/J_{1}& =& \cases{M+1 &
    for $\gamma \geq -1/M$ \cr
    1-1/\gamma & for $\gamma < -1/M$}\label{fm1} \\
  \Omega_{2}(\gamma)/J_{1}& =& \cases{1+M\gamma &
    for $\gamma \geq -1/M$ \cr
    0 & for $\gamma < -1/M$} \label{fm2}\\
  \Omega_{3}(\gamma)/J_{1} &= &\cases{M|1-\gamma| &
    for $1\ne\gamma \geq -1/M$ \cr
    1-1/\gamma & for $\gamma < -1/M$} \label{fm3}\\
  \Omega_{4}(\gamma)/J_{1}& =& \cases{|M(2-\gamma)+1| &
    for $\gamma \geq -1/M$ \cr
    2 \left( 1-1/\gamma \right) & for $\gamma < -1/M$,\label{fm4}}
\nonumber\\
\end{eqnarray}
  and for antiferromagnetic couplings, we find,
\begin{eqnarray}
  \Omega_{1}(\gamma)/|J_{1}| &=& \cases{\left| 1-1/\gamma
    \right| &
    for $\gamma \geq 1/M$ \cr
    M-1 & for $\gamma < 1/M$} \label{afm1}\\
  \Omega_{2}(\gamma)/|J_{1}| &=& \cases{0 &
    for $\gamma \geq 1/M$ \cr
    1-M\gamma & for $\gamma < 1/M$}\label{afm2} \\
  \Omega_{3}(\gamma)/|J_{1}|& =& \cases{\left| 1-1/\gamma
    \right| &
    for $\gamma \geq 1/M$ \cr
    M(1-\gamma) & for $\gamma < 1/M$}\label{afm3} \\
  \Omega_{4}(\gamma)/|J_{1}| &=& \cases{2 \left| 1-1/\gamma
    \right| &
    for $\gamma \geq 1/M$ \cr
    M(2-\gamma)-1 & for $\gamma < 1/M$.\label{afm4}}\nonumber\\
\end{eqnarray}
We remark that these formulae also apply for the isosceles triangle, $M=2$.\cite{isosceles}

\section{V. Low Temperature Numerical Results for $N=4$}

In Fig.  1, we plot the mode frequencies $\Omega_i(\gamma)$
 relative to $|J_1|$, for the squashed tetrahedron case $M=3$.  The upper and
 lower panels correspond to the FM and AFM cases, respectively. The circle in the upper panel of Fig. 1
  denotes the absence of a zero-frequency peak at all temperatures for the regular tetrahedron. We have
  verified these mode frequencies by numerical evaluation of the  explicit
 integral representations of $\delta\tilde{\cal C}_{11}(\omega)$ and
 $\delta\tilde{\cal C}_{44}(\omega)$.  
 For example, in Fig. 2 we show the low-$T$ behavior of $\delta\tilde{\cal
 C}_{11}(\omega)$, presented as $\log_{10}[\delta\tilde{\cal C}_{11}(\omega)]$
 versus $\omega/|J_1|$.  For the FM case with $\gamma=0.3$ at
 $\alpha=50$ pictured in the upper panel of Fig. 2, 
 $\delta\tilde{\cal C}_{11}(\omega)$ exhibits  very sharp peaks at the
 frequencies $\Omega_i$, where $\Omega_i/J_1 = 4, 1.9, 2.1$,
 and 6.1 for $i=1,\ldots,4$, respectively.  $\delta\tilde{\cal
 C}_{44}(\omega)$ has a single sharp  mode at the frequency $\Omega_1$. 
 This figure also shows that for $\gamma=0.6$, the FM $\delta\tilde{\cal
 C}_{11}(\omega)$ modes are also sharp at
 $\alpha=50$, appearing at 4, 2.8, 1.2, and 5.2, respectively, and at
 $\gamma=0.9$, they appear at 4, 3.7, 0.3, and 4.3, respectively.  We note
 that the $\Omega_4$ mode is much weaker in intensity than the other modes at
 this temperature. 
\begin{figure}
\includegraphics[width=0.45\textwidth]{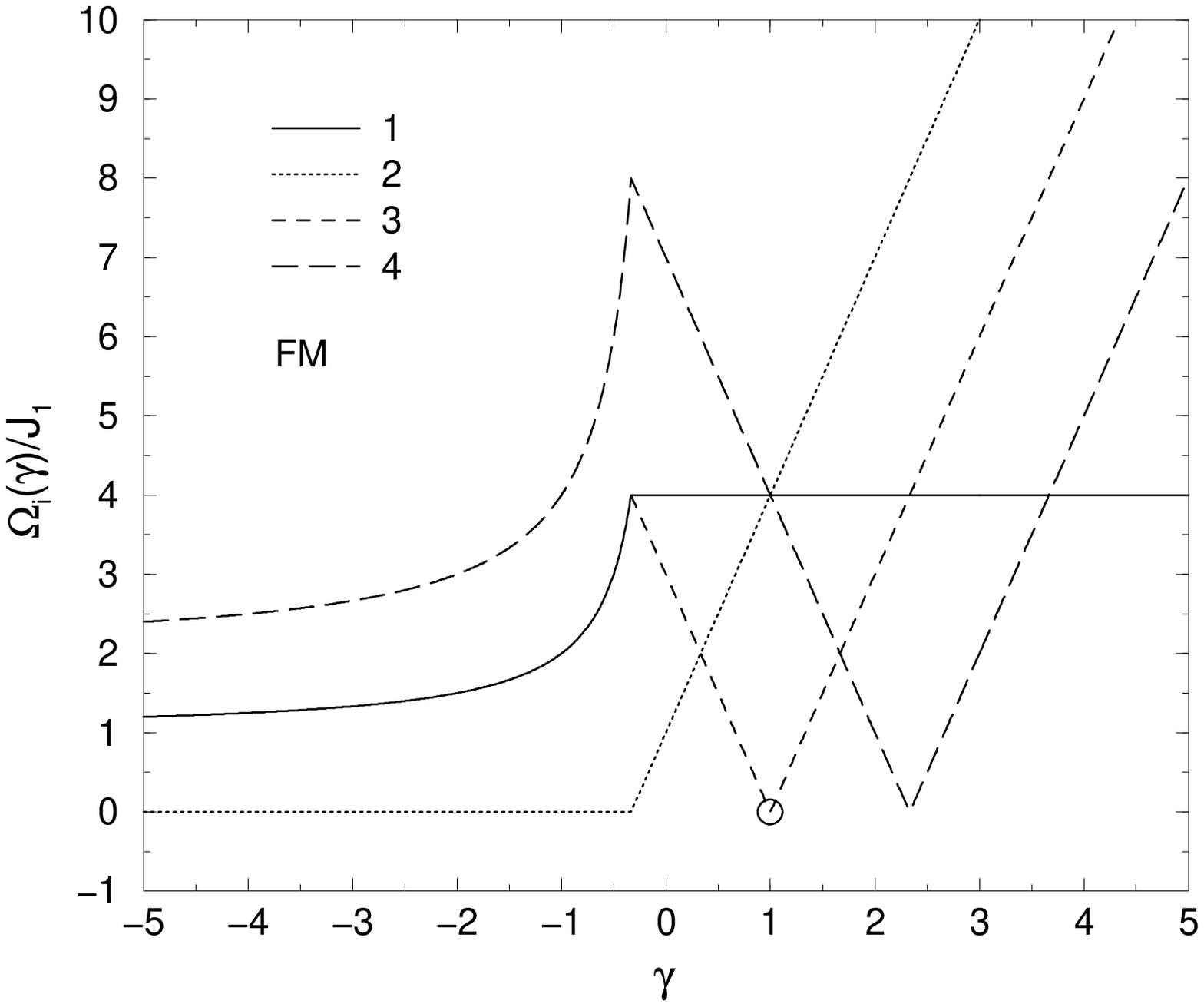}
\includegraphics[width=0.45\textwidth]{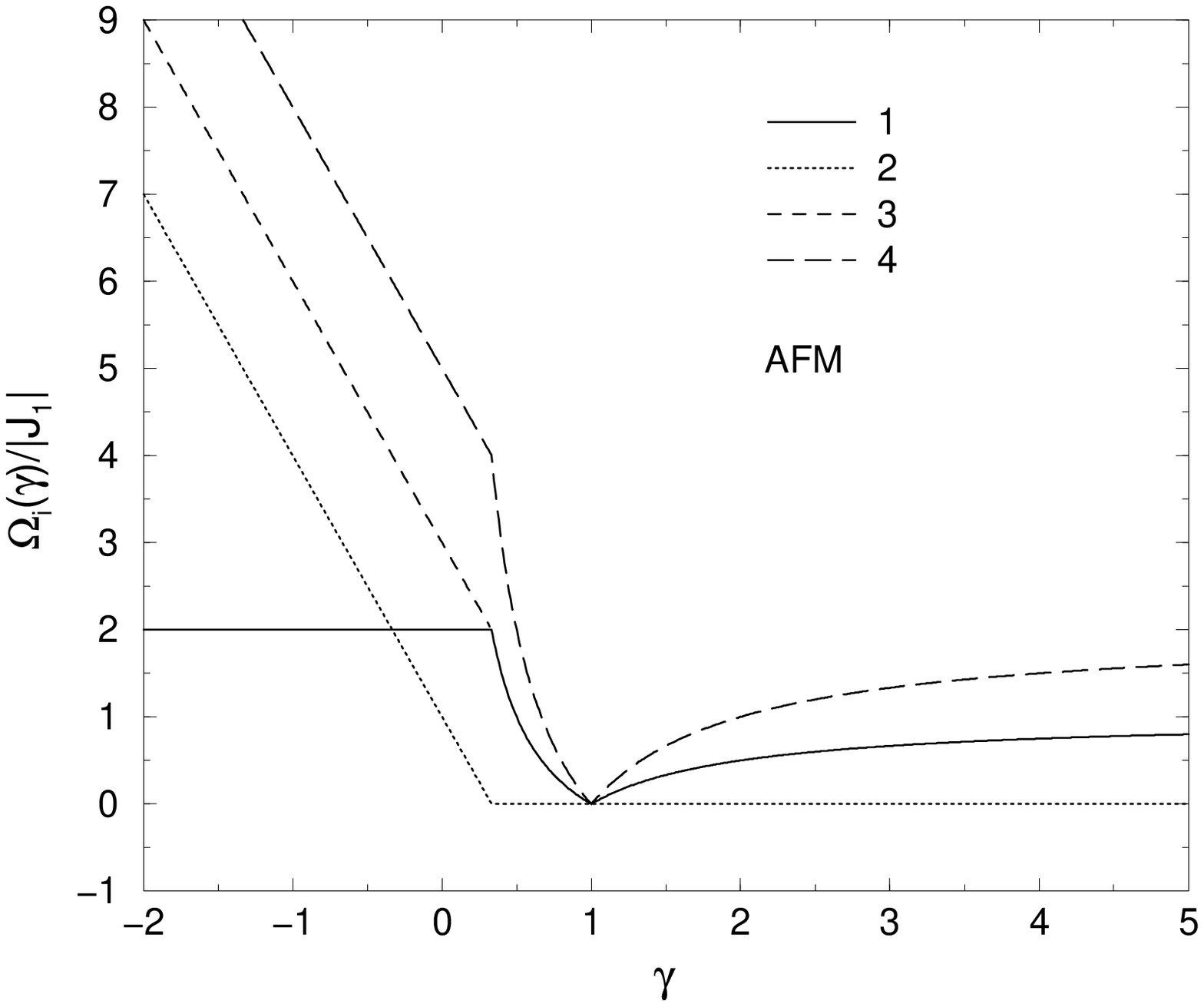}
 \caption{The low-$T$ magnon mode frequencies for the FM (top)
  and AFM (bottom) cases for the squashed tetrahedron ($M=3$).}
 \label{fig1}
\end{figure}
For the AFM case, the modes tend to be much broader, as pictured for
$\gamma=0.6$ in the bottom panel of Fig. 2.  In this case, the low-$T$ mode
frequencies satisfy $\Omega_i/|J_1|=2/3,0,2/3,4/3$, so that $\Omega_1$ and
$\Omega_3$ are degenerate.  This degeneracy is evident in the shape of the
combined mode, which appears to consist of two peaks
with different widths, both centered  at $\omega/|J_1|=2/3$. In addition, $\Omega_2$ is a central peak, which grows in
intensity as $T$ decreases.

\begin{figure}
\includegraphics[width=0.45\textwidth]{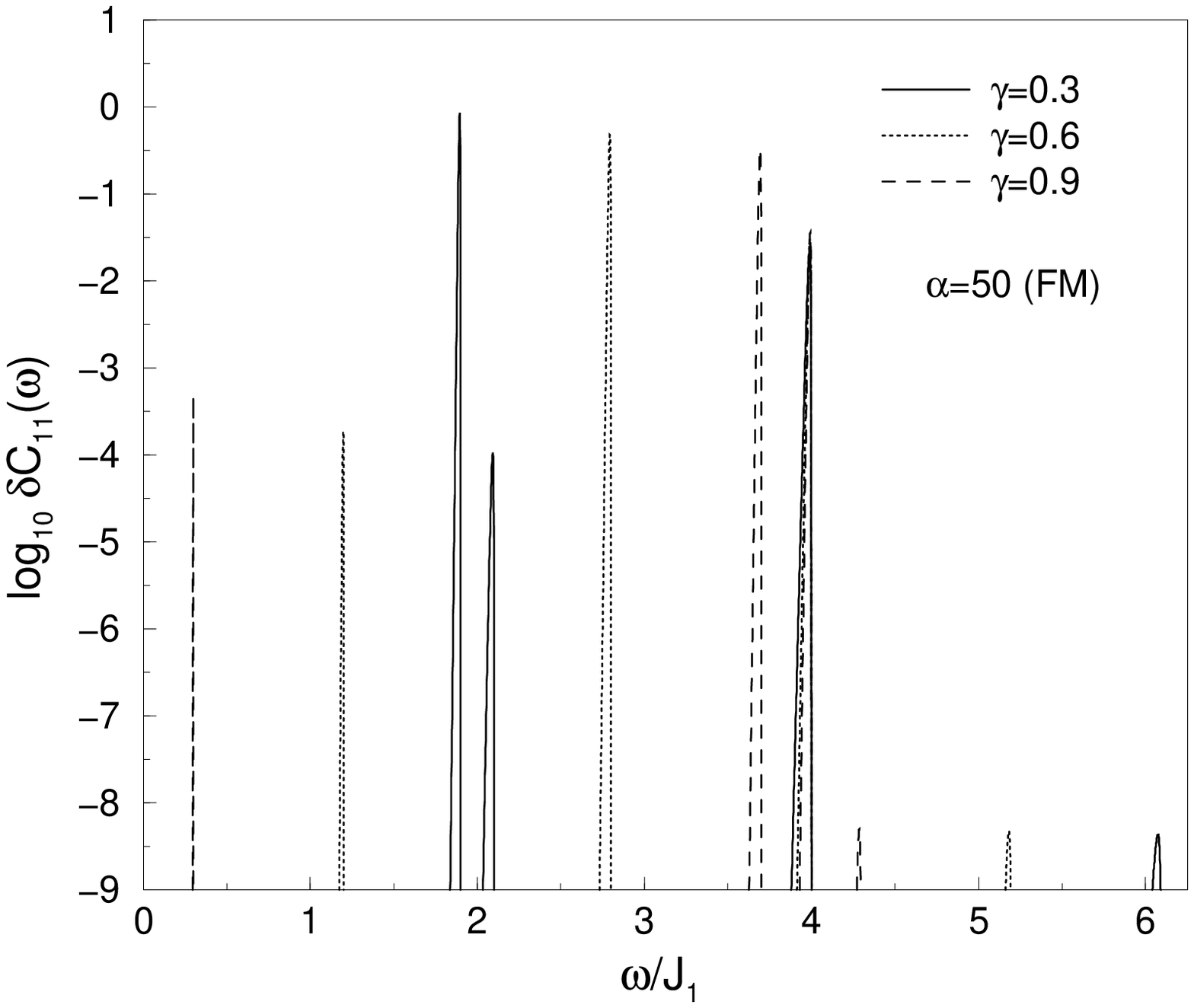}
\includegraphics[width=0.45\textwidth]{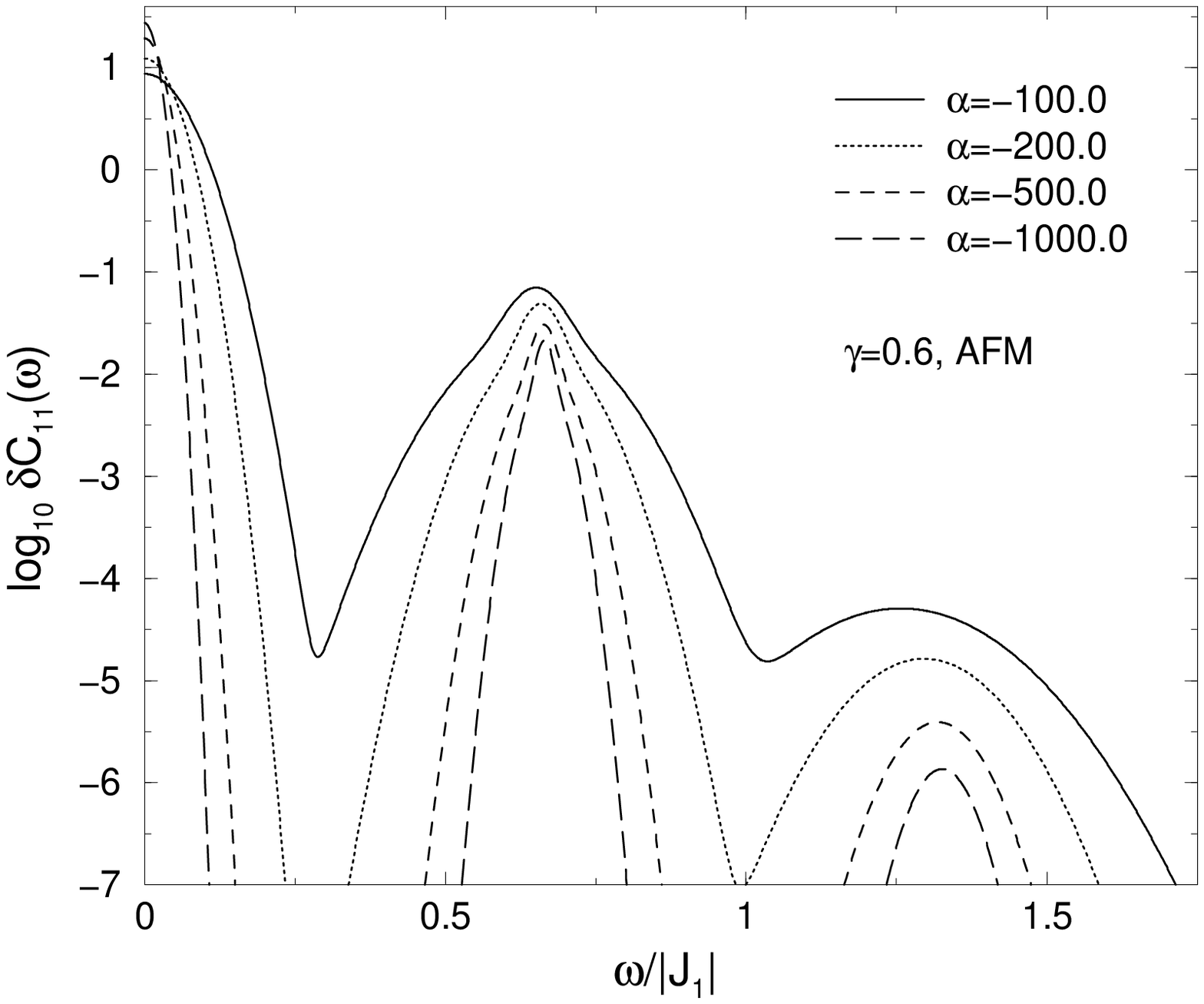}
 \caption{Plots of $\log_{10}[\delta\tilde{\cal C}_{11}(\omega)]$ versus
$\omega/|J_1|$ for the squashed tetrahedron ($M=3$) at very low $T$.  Top:  FM case  at $\alpha=50$ for
$\gamma=0.3, 0.6, 0.9$. Bottom:
   AFM case for $\gamma =0.6$ at various low $T$ values.}
 \label{fig2}
\end{figure}

For the special case of the three-pronged star, $\gamma=0$, the leading
behaviors of the low-$T$ modes are presented in Fig. 3.  For the FM star,
pictured in the upper panel of Fig. 3, we have plotted $n^2\delta\tilde{\cal
C}_{ii}(\omega)$ versus $|\alpha|(\omega/J_1-n)/n$ for the largest amplitude
modes $\Omega_1$ and $\Omega_2$, for $i=1,4$.  Since $\Omega_1$ and $\Omega_2$
appear at $\omega/J_1=4,1$, respectively, and since the $\Omega_1$ modes
present in $\delta\tilde{\cal C}_{44}$ and $\delta\tilde{\cal C}_{11}$ are
weaker than the $\Omega_2$ mode, this presentation was chosen for clarity.
Each of these modes was plotted at $\alpha=5,10,$ and 20, demonstrating the
low-$T$ scaling that occurs.  We also note that the $\Omega_2$ mode in
 $\delta\tilde{\cal C}_{11}(\omega)$ drops discontinuously by many orders of
magnitude (and to zero as $T\rightarrow0$) at $\omega/J_1=1$, as indicated by
the $\approx$ sign.  This behavior is very similar to that of the FM chain,
except for the difference in the frequencies involved.\cite{isosceles}

The AFM three-pronged star has parameters close to those present in the
squashed tetrahedron Fe$_4$.\cite{Fe3tetra:1,Fe3tetra:2}  The strongest low-$T$
modes are pictured in the bottom panel of Fig. 3, in which we plotted
$\delta\tilde{\cal C}_{ii}(\omega)$ versus $|\alpha|(\omega/|J_1|-n)/n$ for
$i=1,4$, $n=1,2$, and $\alpha=-5, -10, -20$, and -40.  Since these modes are
sufficiently close in magnitude, the $\delta\tilde{\cal C}_{ii}(\omega)$ are
 not scaled in this figure.  The additional modes at $\omega/|J_1|=3,5$ are
very weak, and are not shown.  As for the FM case, $\Omega_2$ drops
discontinuously by orders of magnitude at $\omega/|J_1|=1$, vanishing as
$T\rightarrow0$.  In addition, in both cases, the mode shapes approach uniform
functions of $|\alpha|(\omega/|J_1|-n)$ as $T\rightarrow0$.  This behavior is
actually simpler than that obtained for the AFM chain, \cite{isosceles}
because in that case, the $\Omega_1$ and $\Omega_2$ modes present in
$\delta\tilde{\cal C}_{11}(\omega)$ both approach the same frequency,
$\omega/|J_1|=1$, as $T\rightarrow0$, making it difficult to separate them.

 It is interesting to compare these findings with the simpler results
 in the case of a perfect tetrahedron (the equivalent neighbor model
 with $N=4$. \cite{eq_neighbor} There only one low-$T$ mode is
 present, at $\Omega/J=4$ in the ferromagnetic case, or at $\Omega=0$
 in the antiferromagnetic case.  The low-$T$ scaling of these single modes was
 shown previously.\cite{eq_neighbor}  Allowing one spin to be 
 coupled differently
 induces a splitting in the spectrum of low-$T$ magnons, a phenomenon
 which was already observed in the study of the isosceles triangle
 of spins. \cite{isosceles}

\begin{figure}
\includegraphics[width=0.45\textwidth]{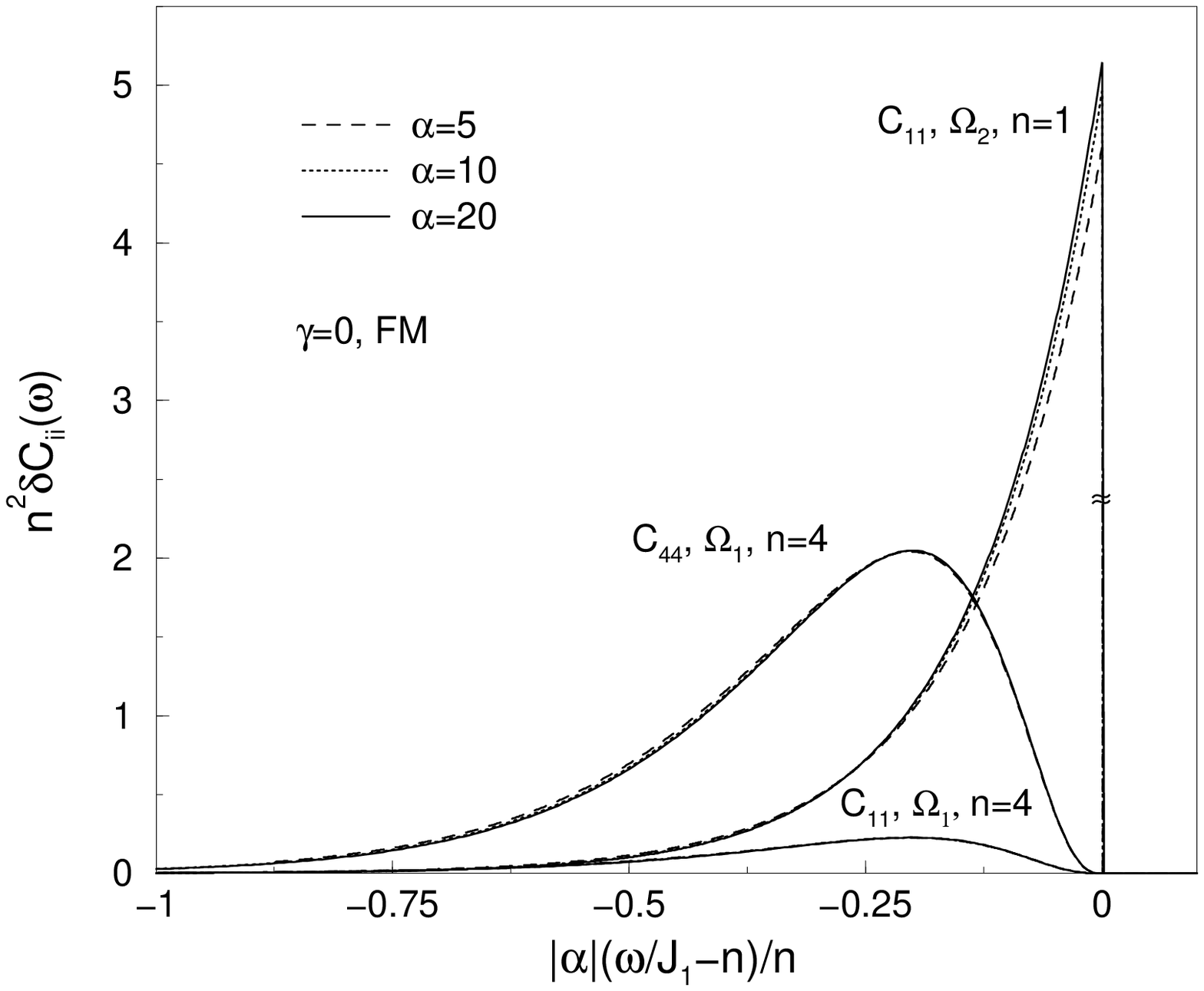}
\includegraphics[width=0.45\textwidth]{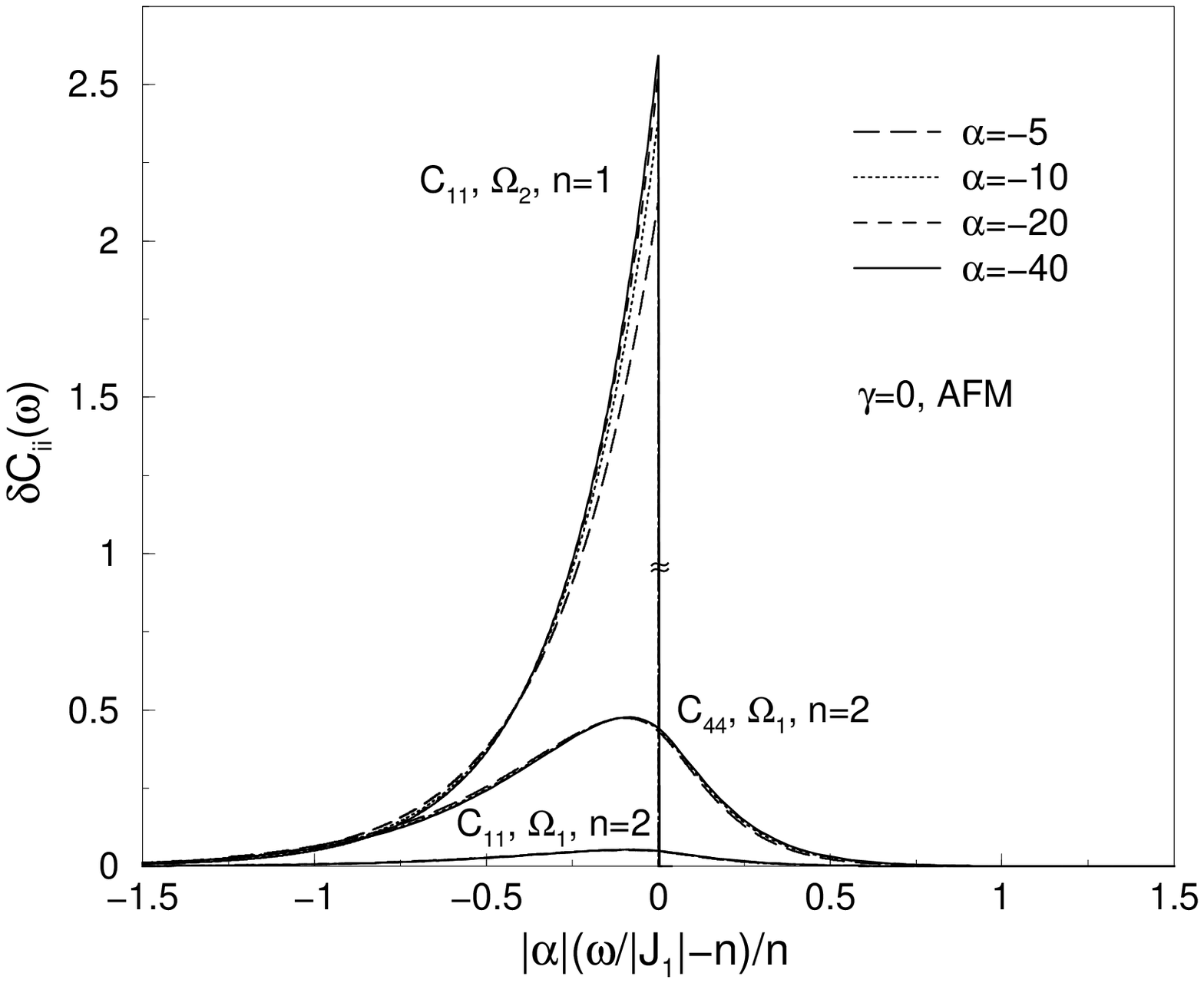}
 \caption{Top: Plots for $M=3$ of the $\Omega_1$ and $\Omega_2$ modes in 
 $n^2\delta\tilde{\cal C}_{ii}(\omega)$ versus
              $|\alpha|(\omega/J_1-n)/n$ with $i,n=1,4$ for the FM star,
             $\gamma=0$, at $\alpha=5,10,20$.
   Bottom: Plots for $M=3$ of the $\Omega_1$ and $\Omega_2$ modes in
 $\delta\tilde{\cal C}_{ii}(\omega)$ versus
             $|\alpha|(\omega/|J_1|-n)/n$ with $i=1,4$ and $n=1,2$, for the
             AFM star, $\gamma=0$, at $\alpha=-5, -10, -20$, and -40.}
\label{fig3} 
\end{figure}

\begin{figure}[floatfix]
\includegraphics[width=0.45\textwidth]{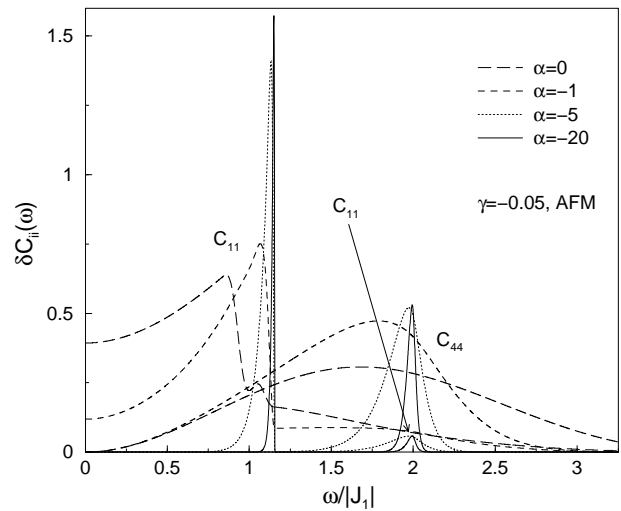}
\caption{ Plots for $M=3$ of $\delta\tilde{\cal C}_{ii}(\omega)$ versus $\omega/|J_1|$ at
various temperatures, for the AFM case with $\gamma=-0.05$, appropriate for 
Fe$_4$.
\cite{Fe3tetra:1}} 
\label{fig4}
\end{figure}

In Fig. 4, we plot the full temperature dependence of the two primary modes
present for the AFM case with $\gamma=-0.05$, which is thought to be a better
approximation to the parameters present in Fe$_4$ than in the bottom panel of
Fig. 3. \cite{Fe3tetra:1,Fe3tetra:2}  In this figure, we show the results of
calculations for both $\delta\tilde{\cal C}_{11}(\omega)$ and
$\delta\tilde{\cal C}_{44}(\omega)$ at $\alpha=0,-1,-5$, and -20.  At
infinite $T$, $\alpha=0$, $\delta\tilde{\cal C}_{44}(\omega)$ exhibits a
broad peak with a maximum at $\omega/|J_1|\approx 1.7$, and
$\delta\tilde{\cal C}_{11}(\omega)$ has substantial weight at low frequencies,
a well-defined peak at $\omega/|J_1|\approx 0.8$, and a small peak at
$\omega/|J_1|\approx1.05$.  As $T$ is lowered, the peak in $\delta\tilde{\cal
C}_{44}(\omega)$ develops into the sharp $\Omega_1$ mode, approaching
$\omega/|J_1|=2$ as $T\rightarrow0$.  In addition, $\delta\tilde{\cal
C}_{11}(\omega)$ develops into the two modes $\Omega_2$ and $\Omega_1$ at
$\omega/|J_1|=1.05$ and 2, respectively.  The minor peaks at
$\omega/|J_1|\approx 3.15$ and 5.15 are too weak to show up on the scale used 
in this figure.  

Finally, in Fig. 5 we show low-$T$ plots at $N=4$ of $\delta\tilde{\cal
C}_{11}(\omega)$ versus $\omega|\alpha|^{1/2}/|J_1|$ for the special points $\gamma=\pm1/3$, corresponding to the
onsets of the central peak of the mode $\Omega_2$.  In both cases, curves for
$|\alpha|=160, 1280$ are shown.  Remarkably, the FM and AFM cases are nearly
identical, when plotted in this manner.  As for the similar scalings at the
endpoints of the the parameter range of the central peak for the isosceles
triangle, \cite{isosceles} this scaling only applies to the frequency, without
a corresponding scaling of $\delta\tilde{\cal C}_{11}(\omega)$, so that the
overall scaling does not correspond to a scaling of the time in $\delta{\cal
C}_{11}(t)$.  However, for the isosceles triangle, the FM and AFM cases
appeared to be nearly similar at temperatures that differed by a factor of
about 8, whereas for the squashed tetrahedron, the temperatures are
essentially identical.

\begin{figure}
\includegraphics[width=0.45\textwidth]{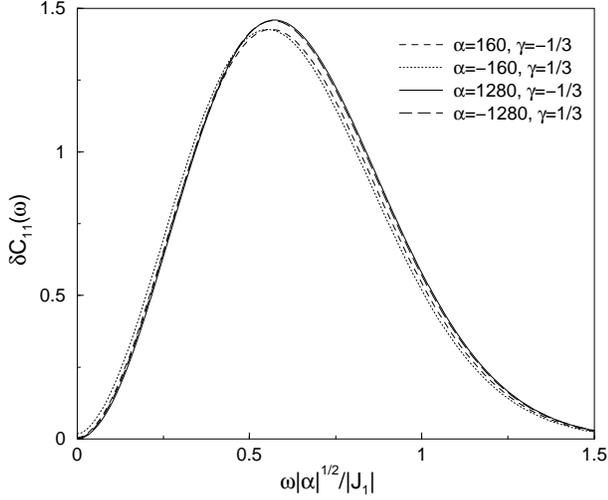}
\caption{ Plots for $M=3$ of $\delta\tilde{\cal C}_{11}(\omega)$ versus
$\omega|\alpha|^{1/2}/|J_1|$ at $|\alpha|=160, 1280$ for the $\Omega_2$ modes
at the onsets of the central peak, $\gamma=\pm1/3$ for the FM and AFM cases, 
respectively.}
\label{fig5}
\end{figure}


\section{VI. Conclusions} 

We have solved for the time correlation functions of the $N$-spin squashed
equivalent neighbor model, with one
spin coupled via the classical Heisenberg exchange $J_1$ to  the $M=N-1$ other spins,
all of which are 
coupled to each other via a different Heisenberg exchange $J_2$.  Our results are qualitatively similar to those of
the isosceles triangle, $N=3$, but show that for arbitrary $N\ge3$, there are
only four low-temperature modes, given by Eqs. (\ref{fm1})-(\ref{fm4}) and
(\ref{afm1})-(\ref{afm4})  for ferromagnetic and antiferromagnetic
signs of $J_1$, respectively.  

At infinite $T$, we showed explicitly that the long-time asymptotic behavior
of the autocorrelation function ${\cal C}_{11}^{\gamma=0}(t)$  on a prong of an $M$-pronged star approaches
its asymtotic limit as $(t^*)^{-1}$.  We also showed that for $3\le N\le8$, the
infinite-$T$, long-time asymptotic limit of ${\cal C}_{NN}(t)$ is greater
than that of ${\cal C}_{11}^{\gamma\ne1}(t)$, and speculate that this relation is likely to
hold for arbitrary $N$.  We also showed that at infinite $T$,  ${\cal
C}_{11}^{0\ne\gamma\ne1}(t)$ approaches its long-time asymptotic limit as
$(\overline{t})^{-M}$, one power slower than does ${\cal C}_{NN}(t)$.

We showed explicitly that these mode
 frequencies apply for the isosceles triangle $(N=3)$ and for the squashed
 tetrahedron ($N=4$). \cite{isosceles}
  For the particular parameter values appropriate for
the single molecule magnet Fe$_4$, with four $S=5/2$ Fe$^{+3}$ spins on the
corners of a squashed tetrahedron, we expect that this classical calculation
of the Fourier transform of the time correlation functions will represent a
reasonably good envelope of the $\delta$-functions present in the quantum
mechanical treatment of this model, provided that the temperatures are not too
low with respect to $|J_1|$.  Thus, we expect the qualitative features shown
in Fig. 4 and the lower panel of Fig. 3 to be observable in inelastic neutron
scattering studies of single crystals of Fe$_4$.


\appendix
\section*{Appendix}
\setcounter{section}{1}

 The integrals appearing for $M\ge3$ in Eq. (\ref{C11}) are
\begin{eqnarray}
  I_{0}& =& \langle S_{1z0}^{2} \rangle ,\label{I0} \\
  I_{1}(t)& =& \langle \frac{A_{N}^{2}S_{1z0}^{2}}{C_{1\rightarrow M}^{2}}
              \cos (st^{*}) \rangle ,\label{I1} \\
  I_{2}(t) &=& \frac{1}{2} \langle (\Delta S_{1z0})^{2}
              \cos [(1-\gamma)xt^{*}] \rangle , \label{I2} \\
  I_{3}(t) &=& \frac{1}{4} \langle A_{N}^{2}(\Delta S_{1z0})^{2}\Bigl(
       \frac{\cos \left\{ [s+(1-\gamma)x]t^{*} \right\}}{(x+C_{1\rightarrow
  M})^{2}}\nonumber\\
& & + \frac{\cos \left\{ [s-(1-\gamma)x]t^{*} \right\}}{(x-C_{1\rightarrow M})^{2}}
       \Bigr) \rangle .\label{I3}
\end{eqnarray}

The density of states for $N$ spins is given by \cite{eq_neighbor}
\begin{eqnarray}
{\cal D}_N(x)&=&\Theta(x)\sum_{p=0}^{E[(N-1)/2]}\Theta(N-2p-x)\nonumber\\
& &\times\Theta(x-N+2p+2)d_{N-2p}(x),\\
d_{N-2p}(x)&=&\sum_{k=0}^p\frac{(-1)^k(N-2k-x)^{N-2}}{2^{N-1}(N-2)!}{N\choose k},\label{DOS}\nonumber\\
\end{eqnarray}
where $E(x)$ is the largest integer in $x$ and $\Theta(x)$ is the
 Heaviside step function.  As noted in Eq. (\ref{Z2}), the $I_i(t)$ contain
 integrations over ${\cal D}_{M-1}(y)$.

Although the $I_i(t)$ are explicitly triple integrals over $x$, $y$, and $s$, the only
$y$ dependence of the integrand appears in the expressions for $S_{1z0}^2$ and
$(\Delta S_{1z0})^2$, given by Eqs. (\ref{S1z0}) and (\ref{DeltaS1z0}) plus
the expressions following Eq. (\ref{SNoft}).  In most of these integrals, one has to evaluate
\begin{eqnarray}
I_i&=&\int_0^{M-1}dy\int_{|y-1|}^{y+1}dx g_M(x,y)f(x,t)\\
&=&\int_0^{M-2}dx f(x,t)\int_{|x-1|}^{x+1}dy g_M(x,y)\nonumber\\
& &+\int_{M-2}^{M}dx f(x,t)
\int_{|x-1|}^{M-1}dy g_M(x,y),
\end{eqnarray}
where $g_M(x,y)$ has either the form $a(x)[1-(x^2+1-y^2)^2/(4x^2)]{\cal
D}_{M-1}(y)$ or the form $a(x)(x^2+1-y^2)^2{\cal D}_{M-1}(y)$, and $f(x,t)$ involves an
integral over $s$.  In most cases, the $y$ integrals can  be performed before the $x$ integrals, reducing the
triple integrals to double integrals, precisely as was done for the equivalent
neighbor model with $M\rightarrow N$.\cite{eq_neighbor}

We now calculate 
 the exact infinite-time, infinite-temperature limit of the
 correlation function  ${\cal C}_{11}(t)$ for $\gamma\ne1$ 
 from Eq. (\ref{I0}).  We first perform the integration over $s$, and then 
invert the order of the remaining two integrations, as
 outlined above.  For $N=4$, we then find,
\begin{eqnarray} \label{exact_limit}
  \lim_{{t\rightarrow \infty}\atop{T\rightarrow \infty}}
   {\cal C}_{11}^{\gamma\ne1}(t)& =& \frac{29}{360} + \frac{\pi^{2}}{384}
   + \frac{83}{360}\ln 2+ \frac{3}{40} \ln 3 \nonumber\\
& &  - \frac{1}{96} \left[ {\rm Li}_{2}\left( -\frac{1}{2} \right)
   + {\rm Li}_{2}\left( -\frac{1}{3} \right) \right]\nonumber\\
& &- \frac{1}{192}\left[ \ln \left(
   \frac{2}{3} \right) \right]^{2}\\
&\approx&0.355496,\nonumber
\end{eqnarray}
 where ${\rm Li}_{2}(z)$ is the standard dilogarithm function,
\begin{equation}
 {\rm Li}_{2}(z) = \int_{z}^{0} \frac{\ln(1-t)}{t}dt.
\end{equation}
 The exact formulae become increasingly
complicated with increasing $N$, so in Table I, we only list the numerical values of those
additional ones for $3\le N\le 11$, along with those of the infinite $t,T$
limits of ${\cal C}_{NN}(t)$.  

\begin{table}
\begin{tabular}{l c c }
\hline
\noalign{\vspace{10pt}}
N & $\>\>\lim_{{t\rightarrow\infty}\atop{T\rightarrow\infty}}
{\cal C}_{11}^{\gamma\ne1}(t)\>\>$ & 
$\>\>\lim_{{t\rightarrow\infty}\atop{T\rightarrow\infty}}
{\cal C}_{NN}(t)\>\>$\\
\noalign{\vspace{10pt}}
\hline
3& 0.370130 & 0.480521\\
4& 0.355496 & 0.436345\\
5& 0.342702 & 0.416362\\
6& 0.337024 & 0.401888\\
7& 0.333611 & 0.384419\\
8& 0.331595 & 0.378635\\
9& 0.330327 & 0.374027\\
10&0.329516 & 0.370270\\
11&0.328992 & 0.367148\\
\hline
\end{tabular} 
\caption{Infinite $t,T$ limits of the autocorrelation functions}
\end{table}

Next, we sketch our procedure for obtaining $ \lim_{T\rightarrow \infty}
   \delta{\cal C}_{11}^{0\ne\gamma\ne1}(t)$  as $\overline{t}\gg1$ for arbitrary $N$.  From
   Eq. (\ref{I2}), we first perform the integration over $s$, giving us a
   function proportional to $f(x)$ given by Eq. (\ref{fofy}).  
To avoid the singularity at $x=1$,
   we do not invert the order of the remaining two integrals, but instead
   integrate with respect to $x$ by parts twice, leading to 
\begin{eqnarray}
\lim_{{T \rightarrow \infty} \atop{\overline{t} \gg 1}} \delta 
     {\cal C}_{11}^{\gamma \neq 0,1}(t) &\sim
     &-\frac{1}{8\overline{t}^2}\sum_{\sigma=\pm1}\int_0^{M-1}ydy{\cal D}_{M-1}(y)\nonumber\\
& &\times f(y+\sigma)\cos[(y+\sigma)\overline{t}].
\label{C11asymp}
\end{eqnarray}
We then integrate with respect to $y$ a total of $M-2$ times, noting 
that all terms
proportional to derivatives of $f$ sum to zero.  We finally obtain,
\begin{eqnarray}
\lim_{{T \rightarrow \infty} \atop{\overline{t} \gg 1}} \delta 
     {\cal C}_{11}^{\gamma \neq 0,1}(t) &\sim
     &\frac{1}{(\overline{t})^M}\sum_{p=0}^{{\rm E}(M/2)}A_{Mp}
f(M-2p)\nonumber\\
& &\times\cos[(M-2p)\overline{t}+M\pi/2],\>\>\>\\
A_{Mp}&=&\frac{(-1)^{p+M}}{2^{M+1}}\Bigl[(1-\delta_{p,M/2})\nonumber\\
& &\times(M-2p-1){M-1\choose p}\nonumber\\
& &-(1-\delta_{p,0})(M-2p+1)\nonumber\\
& &\times{M-1\choose p-1}\Bigr],
\end{eqnarray}
where $f(x)$ is given by Eq. (\ref{fofy}).
 We note that $f(0)=\frac{8}{3}$
and $f(1)=2$. 

We now sketch our derivations of the low-temperature mode frequencies.  We
first note from Eqs. (\ref{CNN}) and (\ref{I1}) that the Fourier transforms of
$\delta{\cal C}_{NN}(t)$ and $I_1(t)$ both contain $\delta(s-\tilde{\omega})$,
where $\tilde{\omega}=\omega/|J_1|$.  From the above discussion, each of these
then can be reduced to a single integral over $x$,
\begin{equation}
K_0(\tilde{\omega})=\int_{|\tilde{\omega}-1|}^{\rm
min(M,\tilde{\omega}+1)}dxQ_N(x,\tilde{\omega})e^{[\alpha(\gamma-1)x^2+\tilde{\omega}^2]},
\end{equation}
where $Q_N(x,\tilde{\omega})$ is different for $\delta\tilde{\cal C}_{NN}(\omega)$ and the
$\Omega_1$ mode contribution to $\delta\tilde{\cal C}_{11}(\omega)$. In both
cases it is independent of $\alpha$ and $T$, and is therefore irrelevant to
the determination of the mode frequency $\Omega_1$ in the limit
$T\rightarrow0$.  The integration limits arise from the condition that the
$\delta$-function is restricted by $|x-1|\le s\le x+1$.  For the FM case, $\alpha>0$, we first consider the case
$\gamma<0$.  As $\alpha\rightarrow\infty$, the integral is maximized by
choosing $x$ to have its minimum value, $x=|\tilde{\omega}-1|$.  We then
maximize the resulting expression for the exponent as a function of
$\tilde{\omega}$, which occurs at $\tilde\omega=\tilde{\omega}^*=1-1/\gamma$.
For $\gamma>1$, the minimum $x$ value, $|\tilde{\omega}-1|$, is limited for
large $\tilde{\omega}$ by $M$, so $\tilde\omega^*=N$. The crossover occurs when
these frequencies are equal,  $N=1-1/\gamma$, or $\gamma=-1/M$.  Setting
$\tilde{\omega}^*=\Omega_1/|J_1|$, we thus recover Eq. (\ref{fm1}).  For the AFM
case as $T\rightarrow0$, $\alpha\rightarrow-\infty$, we want to minimize
$(\gamma-1)x^2+\tilde{\omega}^2$ in the exponent.  For $\gamma>1$, this occurs at
$x=|\tilde{\omega}-1|$, and for $\gamma<1$, it occurs at $x=\tilde\omega+1$.  In
both cases, optimizing the exponent leads to $\tilde\omega^*=|1-1/\gamma|$.
The latter case is restricted by the limitation  
$\tilde{\omega}^*=M-1$.  The crossover between these two limits occurs at
$M=|1-1/\gamma|$, or $\gamma=1/M$.  Setting $\tilde{\omega}^*=\Omega_1/|J_1|$,
we then recover Eq. (\ref{afm1}). 

We now focus on the integral $I_2(t)$, Eq. (\ref{I2}). We first perform the $y$
 integral as sketched above. Then, the integral over $s$ does not contain any
 time dependence, and as
 $T\rightarrow0$, it is dominated by the factor $\exp(\alpha s^2)$.  After
 integration by parts, we obtain the single integral over $x$, which has the
 form
\begin{equation}
I_2(t)\sim\int_0^{M}dx P_N(x)\exp[\alpha(\gamma x^2\pm2x)]\cos[(1-\gamma)xt^*],
\end{equation}
where $P_N(x)$ is independent of $\alpha$, as in Eq. (\ref{Z}). Fourier
 transformation then
  involves the $\delta$-function,
 $\delta(\tilde{\omega}-|1-\gamma|x)$, so that the position of the mode due to
 $I_2$ is found by optimizing the expression
 $\exp\{\alpha[\gamma\tilde{\omega}^2/(1-\gamma)^2\pm2\tilde{\omega}/|1-\gamma|]\}$.
 For the FM case and $\gamma<0$, we maximize this 
function with the + sign, leading to $\tilde{\omega}^*=1-1/\gamma$. For
 $\gamma>0$, the $\delta$-function was restricted by $x\le M$, leading to
 $\tilde{\omega}^*=M|1-\gamma|$. These values for $\Omega_3/|J_1|=\tilde{\omega}^*$
 are equal at  $\gamma=-1/M$.  Combining, we obtain the FM $\Omega_3$ mode
 frequencies, Eq. (\ref{fm3}). For the AFM case as $\alpha\rightarrow-\infty$,
 we choose the $-$ sign in the above exponent, and minimize
 $\gamma\tilde{\omega}^2/(1-\gamma)^2-2\tilde{\omega}$ in the exponent.  For $\gamma>0$, this
 occurs at $\tilde\omega^*=|1-1/\gamma|$. For $\gamma<0$, the overall exponent
 is bounded by $\tilde{\omega}^*/|1-\gamma|\le M$.  Combining, we obtain the
 expressions for $\Omega_3/|J_1|$ for the AFM case, Eq. (\ref{afm3}).

 We now turn our attention to $I_3$. In taking the Fourier transform, there
 are four $\delta$-functions, $\delta[\tilde{\omega}-s-(1-\gamma)x]$,
 $\delta[\tilde{\omega}+s+(1-\gamma)x]$,
 $\delta[\tilde{\omega}+s-(1-\gamma)x]$, and
 $\delta[\tilde{\omega}-s+(1-\gamma)x]$. These $\delta$-functions lead
 after the usual reductions of the $y$ integrals to the following integrals,
 respectively,
\begin{eqnarray}
K_1(\omega)&=&\int_{{\rm
max}[(1-\tilde{\omega})/\gamma,(\tilde{\omega}-1)/(2-\gamma)]}^{{\rm
min}[M,(\tilde{\omega}+1)/(2-\gamma)]}dx
R_{N1}(x,\tilde{\omega})\nonumber\\
& &\times f_{+}(x,\tilde{\omega}),\label{K1}\\
K_2(\omega)&=&\Theta(\gamma-1)\int_{{\rm
max}[0,(1+\tilde{\omega})/\gamma,(\tilde{\omega}-1)/(\gamma-2)]}^{{\rm
min}[M,\Theta(\gamma-2)(\tilde{\omega}+1)/(\gamma-2)]}dx\nonumber\\
& &\times
R_{N2}(x,\tilde{\omega})f_{-}(x,\tilde{\omega}),\label{K2}\\
K_3(\omega)&=&\Theta(1-\gamma)\int_{{\rm
max}[0,-(1+\tilde{\omega})/\gamma,(\tilde{\omega}+1)/(2-\gamma)]}^{{\rm
min}[M,(1-\tilde{\omega})/\gamma]}dx\nonumber\\
& &\times 
R_{N3}(x,\tilde{\omega})f_{+}(x,\tilde{\omega}),\label{K3}\\
K_4(\omega)&=&\int_{{\rm max}[0,(\tilde{\omega}-1)/\gamma]}^{{\rm
min}[M,(\tilde{\omega}+1)/\gamma]}dx\nonumber\\
& &\times  R_{N4}(x,\tilde{\omega})f_{-}(x,\tilde{\omega}),\label{K4}\\
f_{\pm}(x,\tilde{\omega})&=&
\exp\{\alpha[\tilde{\omega}^2+\gamma(\gamma-1)x^2\pm2\tilde{\omega}(\gamma-1)x]\},\label{fpm}\nonumber\\
\end{eqnarray}
where the $R_{Ni}(x,\tilde{\omega})$ are independent of $T$.

We first consider the AFM case of $K_1$, $\alpha\rightarrow-\infty$.  For
$\gamma>1$, all terms in the exponent are negative, so we need to minimize the
function $\tilde{\omega}^2+\gamma(\gamma-1)x^2+2x\tilde{\omega}(\gamma-1)$.
Setting $x=(\tilde{\omega}-1)/(2-\gamma)$, and optimizing this function with
respect to $\tilde{\omega}$, we find that its minimum occurs at
 $\tilde{\omega}^*=2(1-1/\gamma)$.  For $\gamma<0$, the last term in the
function to be minimized is negative, so we take
 $x=(\tilde{\omega}+1)/(2-\gamma)$ from the upper integration limit.
Optimizing the function, we find $\tilde{\omega}^*=2(1/\gamma-1)$, so both 
 $\gamma$ regions satisfy $\tilde{\omega}^*=2|1-1/\gamma|$.   However, this
is subject to the constraint on the upper integration cutoff, which is
 $(\tilde{\omega}^*+1)/(2-\gamma)=M$, or
 $\tilde{\omega}^*=2M-1-M\gamma$.  These values are equal at
 $\gamma=1/M$.  Altogether, $\tilde{\omega}^*=\Omega_4/|J_1|$ for the AFM
case in Eq. (\ref{afm4}).  For the FM case with $\gamma<0$, we take
$x=(\tilde{\omega}+1)/(2-\gamma)$, optimize, and again obtain
$\tilde{\omega}^*=2(1-1/\gamma)$.  The cutoff occurs when the lower limit,
$x=(\tilde{\omega}-1)/(2-\gamma)$, equals $M$, giving
$\tilde{\omega}^*=2M+1-M\gamma$.  The crossover occurs at
$\gamma=1/M$, as given by Eq. (\ref{fm4}) for $\Omega_4/|J_1|$.

Next, we consider the FM case of $K_4$.  First for
$\alpha\rightarrow\infty$, $\gamma<0$, it is easily seen that the exponent in
$f_{-}(x\tilde{\omega})$ is positive definite.  Thus, we might expect the
upper limit for $x$ to apply.  But, this is either the cutoff, $M$, or a
negative quantity, $(\tilde{\omega}+1)/\gamma$.  Thus, the only positive limit
is the lower cutoff, $x=(\tilde{\omega}-1)/\gamma$, which can be positive for
$\tilde{\omega}<1$, leading to a larger exponent than obtained by setting $x=0$.  However,
$f_{-}[(\tilde{\omega}-1)/\gamma,\tilde{\omega}]=\exp[(\alpha/\gamma)(\tilde{\omega}^2+\gamma-1)]$,
which for $\alpha>0$, $\gamma<0$ has a maximum at $\tilde{\omega}^*=0$,
corresponding to a
central peak.  This will be the mode frequency until
$(\tilde{\omega}-1)/\gamma=M$, the upper cutoff, resulting in
$\tilde{\omega}^*=1+M\gamma$.  The crossover occurs at $\gamma=-1/M$.  Thus, this mode reduces to $\Omega_2/|J_1|$
as given by Eq. (\ref{fm2}).  For the AFM case for $\gamma>0$, we set
$x=(\omega+1)/\gamma$, and again we find
$f_{-}[(\tilde{\omega}+1)/\gamma,\tilde{\omega}]=\exp[-(|\alpha|/\gamma)(\tilde{\omega}^2+\gamma-1)]$,
which has a maximum at $\tilde{\omega}^*=0$.  This form continues until
$(1-\tilde{\omega})/\gamma=M$, which occurs at
$\tilde{\omega}^*=1-M\gamma$.  The crossover occurs at $\gamma=1/M$.  Thus, this gives rise to the mode
$\Omega_2/|J_1|$ in Eq. (\ref{afm2}).

We now consider the $K_2$ integral.  This makes a very small contribution,
because of the the severe limitation that it vanishes unless $\gamma>1$.  For
the AFM case, the exponent is optimized at $x=x^*=\tilde{\omega}/\gamma$, and
then optimizing the mode frequency with respect to $\tilde{\omega}$, we find
that $\tilde{\omega}^*=0$, so that $K_2$ for AFM coupling contributes to
$\Omega_2/|J_1|$.  For the FM case, the maximum exponent occurs at $x=M$,
and from the $\delta$-function restrictions, we see that $K_2$ makes a
contribution to the $\Omega_4/|J_1|$ mode.

Finally, we discuss briefly the $K_3$ case, for which $\gamma<1$.  Setting
$\gamma<0$ for the FM case, the optimum situation is obtained when
$\tilde{\omega}^*=0$, so that it adds to the $\Omega_2/|J_1|$ mode.  For the
AFM case, the optimum $x$ value is $x^*=-\tilde{\omega}/\gamma$, and this
is restricted by $x\le M$.  Hence, $K_3$ essentially makes a contribution to
the $\Omega_2$ mode for the AMF case, as well.



\end{document}